\documentclass[aps,prl,reprint,superscriptaddress,floatfix,nofootinbib,showkeys,showpacs,preprintnumbers]{revtex4-1}

\pdfoutput=1

\usepackage{tabularx}
\usepackage{xspace}
\usepackage{listings}
\usepackage{lineno}

\usepackage[dvipsnames,svgnames]{xcolor} 
\usepackage{graphicx}
\usepackage{comment}

\usepackage{hyperref}
\hypersetup{    
  colorlinks      = true,
  linkcolor       = {blue},
  linkbordercolor = {white},
  citecolor       = {blue},
  citebordercolor = {white},
  urlcolor        = {blue},
  urlbordercolor  = {white},
}

\usepackage{natbib}
\usepackage{hyperref}
\usepackage{amsmath}
\usepackage{enumitem}
\setlist{wide, labelwidth=!, labelindent=0pt}
\usepackage{multirow}

\usepackage{soul} 
\usepackage{amsmath}
\usepackage{amssymb}
\usepackage{xspace}
\usepackage{xifthen}

\newcommand{\CHECK}[1]{{#1}}

\mathchardef\mhyphen="2D
\newcommand{\vect}[1]{\boldsymbol{#1}}
\newcommand{\roughly}{\ensuremath{ {\sim}\,} }

\newlength{\dhatheight}

\newcommand{\unit}[1]{\ensuremath{\mathrm{\,#1}}\xspace}

\newcommand{\eV}{\unit{eV}}
\newcommand{\keV}{\unit{keV}}
\newcommand{\MeV}{\unit{MeV}}
\newcommand{\GeV}{\unit{GeV}}

\newcommand{\cm}{\unit{cm}}

\newcommand{\pc}{\unit{pc}}
\newcommand{\kpc}{\unit{kpc}}
\newcommand{\Mpc}{\unit{Mpc}}

\newcommand{\Msun}{\unit{M_\odot}}

\newcommand{\tabref}[1]{Table~\ref{tab:#1}}

\newcommand{\figref}[1]{Fig.~\ref{fig:#1}}

\newcommand{\eqnref}[1]{Eq.~\eqref{eq:#1}}

\newcommand{\bandvar}[2][]{%
  \ifthenelse{\isempty{#1}}{\var{#2}}{\var{#2\_#1}}%
}

\newcommand{\var}[1]{\ensuremath{\texttt{\MakeUppercase{#1}}}\xspace}

\newcommand{\DM}{\ensuremath{\mathrm{DM}}}
\newcommand{\CDM}{\ensuremath{\mathrm{CDM}}}
\newcommand{\WDM}{\ensuremath{\mathrm{WDM}}}
\newcommand{\FDM}{\ensuremath{\mathrm{FDM}}}

\newcommand{\mWDM}{\ensuremath{m_\WDM}\xspace}
\newcommand{\mFDM}{\ensuremath{m_\phi}\xspace}

\newcommand{\Mhm}{\ensuremath{M_{\rm hm}}\xspace}
\newcommand{\khm}{\ensuremath{k_{\rm hm}}\xspace}

\providecommand\physrep{\ref@jnl{Phys.~Rep.}}%
\providecommand\apjs{\ref@jnl{ApJS}}%
\providecommand{\jcap}{\ref@jnl{JCAP}}%

\usepackage{lipsum}
\usepackage{aas_macros}

\graphicspath{{./}{./figures/}{./tex/figures}}

\begin{document}

\preprint{DES-2020-546}
\preprint{FERMILAB-PUB-20-277-AE}
\preprint{SLAC-PUB-17554}

\title{Milky Way Satellite Census. III. Constraints on Dark Matter Properties from Observations of Milky Way Satellite Galaxies}


\author{E.~O.~Nadler}
\email[]{enadler\@stanford.edu}
\affiliation{Department of Physics, Stanford University, 382 Via Pueblo Mall, Stanford, California 94305, USA}
\affiliation{Kavli Institute for Particle Astrophysics and Cosmology, P. O. Box 2450, Stanford University, Stanford, California 94305, USA}
\affiliation{SLAC National Accelerator Laboratory, Menlo Park, California 94025, USA}
\author{A.~Drlica-Wagner}
\email[]{kadrlica@fnal.gov}
\affiliation{Fermi National Accelerator Laboratory, P. O. Box 500, Batavia, Illinois 60510, USA}
\affiliation{Kavli Institute for Cosmological Physics, University of Chicago, Chicago, Illinois 60637, USA}
\affiliation{Department of Astronomy and Astrophysics, University of Chicago, Chicago, Illinois 60637, USA}
\author{K.~Bechtol}
\affiliation{Physics Department, 2320 Chamberlin Hall, University of Wisconsin--Madison, 1150 University Avenue, Madison, Wisconsin 53706-1390, USA}
\author{S.~Mau}
\affiliation{Department of Physics, Stanford University, 382 Via Pueblo Mall, Stanford, California 94305, USA}
\affiliation{Kavli Institute for Particle Astrophysics and Cosmology, P. O. Box 2450, Stanford University, Stanford, California 94305, USA}
\author{R.~H.~Wechsler}
\affiliation{Department of Physics, Stanford University, 382 Via Pueblo Mall, Stanford, California 94305, USA}
\affiliation{Kavli Institute for Particle Astrophysics and Cosmology, P. O. Box 2450, Stanford University, Stanford, California 94305, USA}
\affiliation{SLAC National Accelerator Laboratory, Menlo Park, California 94025, USA}
\author{V.~Gluscevic}
\affiliation{University of Southern California. Department of Physics and Astronomy, 825 Bloom Walk ACB 439, Los Angeles, California 90089-0484, USA}
\author{K.~Boddy}
\affiliation{Theory Group, Department of Physics, The University of Texas at Austin, Austin, Texas 78712, USA}
\author{A.~B.~Pace}
\affiliation{Department of Physics, Carnegie Mellon University, Pittsburgh, Pennsylvania 15312, USA}
\author{T.~S.~Li}
\altaffiliation{NHFP Einstein Fellow}
\affiliation{Department of Astrophysical Sciences, Princeton University, Peyton Hall, Princeton, New Jersey 08544, USA}
\affiliation{Observatories of the Carnegie Institution for Science, 813 Santa Barbara Street, Pasadena, California 91101, USA}
\author{M.~McNanna}
\affiliation{Physics Department, 2320 Chamberlin Hall, University of Wisconsin--Madison, 1150 University Avenue, Madison, Wisconsin 53706-1390, USA}
\author{A.~H.~Riley}
\affiliation{George P. and Cynthia Woods Mitchell Institute for Fundamental Physics and Astronomy, and Department of Physics and Astronomy, Texas A\&M University, College Station, Texas 77843,  USA}
\author{J.~Garc\'ia-Bellido}
\affiliation{Instituto de Fisica Teorica UAM/CSIC, Universidad Autonoma de Madrid, 28049 Madrid, Spain}
\author{Y.-Y.~Mao}
\altaffiliation{NHFP Einstein Fellow}
\affiliation{Department of Physics and Astronomy, Rutgers, The State University of New Jersey, Piscataway, New Jersey 08854, USA}
\author{G.~Green}
\affiliation{Max Planck Institute for Astronomy, K\"onigstuhl 17 D-69117, Heidelberg, Germany}
\author{D.~L.~Burke}
\affiliation{Kavli Institute for Particle Astrophysics and Cosmology, P. O. Box 2450, Stanford University, Stanford, California 94305, USA}
\affiliation{SLAC National Accelerator Laboratory, Menlo Park, California 94025, USA}
\author{A.~Peter}
\affiliation{Department of Physics, The Ohio State University, Columbus, Ohio 43210, USA}
\affiliation{Department of Astronomy, The Ohio State University, Columbus, Ohio 43210, USA}
\affiliation{Center for Cosmology and Astro-Particle Physics, The Ohio State University, Columbus, Ohio 43210, USA}
\author{B.~Jain}
\affiliation{Department of Physics and Astronomy, University of Pennsylvania, Philadelphia, Pennsylvania 19104, USA}
\author{T.~M.~C.~Abbott}
\affiliation{Cerro Tololo Inter-American Observatory, NSF's National Optical-Infrared Astronomy Research Laboratory, Casilla 603, La Serena, Chile}
\author{M.~Aguena}
\affiliation{Departamento de F\'isica Matem\'atica, Instituto de F\'isica, Universidade de S\~ao Paulo, CP 66318, S\~ao Paulo, SP, 05314-970, Brazil}
\affiliation{Laborat\'orio Interinstitucional de e-Astronomia - LIneA, Rua Gal. Jos\'e Cristino 77, Rio de Janeiro, RJ - 20921-400, Brazil}
\author{S.~Allam}
\affiliation{Fermi National Accelerator Laboratory, P. O. Box 500, Batavia, Illinois 60510, USA}
\author{J.~Annis}
\affiliation{Fermi National Accelerator Laboratory, P. O. Box 500, Batavia, Illinois 60510, USA}
\author{S.~Avila}
\affiliation{Instituto de Fisica Teorica UAM/CSIC, Universidad Autonoma de Madrid, 28049 Madrid, Spain}
\author{D.~Brooks}
\affiliation{Department of Physics and Astronomy, University College London, Gower Street, London, WC1E 6BT, United Kingdom}
\author{M.~Carrasco~Kind}
\affiliation{Department of Astronomy, University of Illinois at Urbana-Champaign, 1002 West Green Street, Urbana, Illinois 61801, USA}
\affiliation{National Center for Supercomputing Applications, 1205 West Clark Street, Urbana, Illinois 61801, USA}
\author{J.~Carretero}
\affiliation{Institut de F\'{\i}sica d'Altes Energies (IFAE), The Barcelona Institute of Science and Technology, Campus UAB, 08193 Bellaterra (Barcelona), Spain}
\author{M.~Costanzi}
\affiliation{INAF-Osservatorio Astronomico di Trieste, via G. B. Tiepolo 11, I-34143 Trieste, Italy}
\affiliation{Institute for Fundamental Physics of the Universe, Via Beirut 2, 34014 Trieste, Italy}
\author{L.~N.~da Costa}
\affiliation{Laborat\'orio Interinstitucional de e-Astronomia - LIneA, Rua Gal. Jos\'e Cristino 77, Rio de Janeiro, RJ - 20921-400, Brazil}
\affiliation{Observat\'orio Nacional, Rua Gal. Jos\'e Cristino 77, Rio de Janeiro, RJ - 20921-400, Brazil}
\author{J.~De~Vicente}
\affiliation{Centro de Investigaciones Energ\'eticas, Medioambientales y Tecnol\'ogicas (CIEMAT), Madrid, Spain}
\author{S.~Desai}
\affiliation{Department of Physics, IIT Hyderabad, Kandi, Telangana 502285, India}
\author{H.~T.~Diehl}
\affiliation{Fermi National Accelerator Laboratory, P. O. Box 500, Batavia, Illinois 60510, USA}
\author{P.~Doel}
\affiliation{Department of Physics and Astronomy, University College London, Gower Street, London, WC1E 6BT, United Kingdom}
\author{S.~Everett}
\affiliation{Santa Cruz Institute for Particle Physics, Santa Cruz, California 95064, USA}
\author{A.~E.~Evrard}
\affiliation{Department of Astronomy, University of Michigan, Ann Arbor, Michigan 48109, USA}
\affiliation{Department of Physics, University of Michigan, Ann Arbor, Michigan 48109, USA}
\author{B.~Flaugher}
\affiliation{Fermi National Accelerator Laboratory, P. O. Box 500, Batavia, Illinois 60510, USA}
\author{J.~Frieman}
\affiliation{Fermi National Accelerator Laboratory, P. O. Box 500, Batavia, Illinois 60510, USA}
\affiliation{Kavli Institute for Cosmological Physics, University of Chicago, Chicago, Illinois 60637, USA}
\author{D.~W.~Gerdes}
\affiliation{Department of Astronomy, University of Michigan, Ann Arbor, Michigan 48109, USA}
\affiliation{Department of Physics, University of Michigan, Ann Arbor, Michigan 48109, USA}
\author{D.~Gruen}
\affiliation{Department of Physics, Stanford University, 382 Via Pueblo Mall, Stanford, California 94305, USA}
\affiliation{Kavli Institute for Particle Astrophysics and Cosmology, P. O. Box 2450, Stanford University, Stanford, California 94305, USA}
\affiliation{SLAC National Accelerator Laboratory, Menlo Park, California 94025, USA}
\author{R.~A.~Gruendl}
\affiliation{Department of Astronomy, University of Illinois at Urbana-Champaign, 1002 West Green Street, Urbana, Illinois 61801, USA}
\affiliation{National Center for Supercomputing Applications, 1205 West Clark Street, Urbana, Illinois 61801, USA}
\author{J.~Gschwend}
\affiliation{Laborat\'orio Interinstitucional de e-Astronomia - LIneA, Rua Gal. Jos\'e Cristino 77, Rio de Janeiro, RJ - 20921-400, Brazil}
\affiliation{Observat\'orio Nacional, Rua Gal. Jos\'e Cristino 77, Rio de Janeiro, RJ - 20921-400, Brazil}
\author{G.~Gutierrez}
\affiliation{Fermi National Accelerator Laboratory, P. O. Box 500, Batavia, Illinois 60510, USA}
\author{S.~R.~Hinton}
\affiliation{School of Mathematics and Physics, University of Queensland,  Brisbane, Queensland 4072, Australia}
\author{K.~Honscheid}
\affiliation{Center for Cosmology and Astro-Particle Physics, The Ohio State University, Columbus, Ohio 43210, USA}
\affiliation{Department of Physics, The Ohio State University, Columbus, Ohio 43210, USA}
\author{D.~Huterer}
\affiliation{Department of Physics, University of Michigan, Ann Arbor, Michigan 48109, USA}
\author{D.~J.~James}
\affiliation{Center for Astrophysics, Harvard and Smithsonian, 60 Garden Street, Cambridge, Massachusetts 02138, USA}
\author{E.~Krause}
\affiliation{Department of Astronomy/Steward Observatory, University of Arizona, 933 North Cherry Avenue, Tucson, Arizona 85721-0065, USA}
\author{K.~Kuehn}
\affiliation{Australian Astronomical Optics, Macquarie University, North Ryde, New South Wales 2113, Australia}
\affiliation{Lowell Observatory, 1400 Mars Hill Road, Flagstaff, Arizona 86001, USA}
\author{N.~Kuropatkin}
\affiliation{Fermi National Accelerator Laboratory, P. O. Box 500, Batavia, Illinois 60510, USA}
\author{O.~Lahav}
\affiliation{Department of Physics and Astronomy, University College London, Gower Street, London, WC1E 6BT, United Kingdom}
\author{M.~A.~G.~Maia}
\affiliation{Laborat\'orio Interinstitucional de e-Astronomia - LIneA, Rua Gal. Jos\'e Cristino 77, Rio de Janeiro, RJ - 20921-400, Brazil}
\affiliation{Observat\'orio Nacional, Rua Gal. Jos\'e Cristino 77, Rio de Janeiro, RJ - 20921-400, Brazil}
\author{J.~L.~Marshall}
\affiliation{George P. and Cynthia Woods Mitchell Institute for Fundamental Physics and Astronomy, and Department of Physics and Astronomy, Texas A\&M University, College Station, Texas 77843, USA}
\author{F.~Menanteau}
\affiliation{Department of Astronomy, University of Illinois at Urbana-Champaign, 1002 West Green Street, Urbana, Illinois 61801, USA}
\affiliation{National Center for Supercomputing Applications, 1205 West Clark Street, Urbana, Illinois 61801, USA}
\author{R.~Miquel}
\affiliation{Instituci\'o Catalana de Recerca i Estudis Avan\c{c}ats, E-08010 Barcelona, Spain}
\affiliation{Institut de F\'{\i}sica d'Altes Energies (IFAE), The Barcelona Institute of Science and Technology, Campus UAB, 08193 Bellaterra (Barcelona), Spain}
\author{A.~Palmese}
\affiliation{Fermi National Accelerator Laboratory, P. O. Box 500, Batavia, Illinois 60510, USA}
\affiliation{Kavli Institute for Cosmological Physics, University of Chicago, Chicago, Illinois 60637, USA}
\author{F.~Paz-Chinch\'{o}n}
\affiliation{Institute of Astronomy, University of Cambridge, Madingley Road, Cambridge CB3 0HA, United Kingdom}
\affiliation{National Center for Supercomputing Applications, 1205 West Clark Street, Urbana, Illinois 61801, USA}
\author{A.~A.~Plazas}
\affiliation{Department of Astrophysical Sciences, Princeton University, Peyton Hall, Princeton, New Jersey 08544, USA}
\author{A.~K.~Romer}
\affiliation{Department of Physics and Astronomy, Pevensey Building, University of Sussex, Brighton, BN1 9QH, United Kingdom}
\author{E.~Sanchez}
\affiliation{Centro de Investigaciones Energ\'eticas, Medioambientales y Tecnol\'ogicas (CIEMAT), Madrid, Spain}
\author{V.~Scarpine}
\affiliation{Fermi National Accelerator Laboratory, P. O. Box 500, Batavia, Illinois 60510, USA}
\author{S.~Serrano}
\affiliation{Institut d'Estudis Espacials de Catalunya (IEEC), 08034 Barcelona, Spain}
\affiliation{Institute of Space Sciences (ICE, CSIC),  Campus UAB, Carrer de Can Magrans, s/n,  08193 Barcelona, Spain}
\author{I.~Sevilla-Noarbe}
\affiliation{Centro de Investigaciones Energ\'eticas, Medioambientales y Tecnol\'ogicas (CIEMAT), Madrid, Spain}
\author{M.~Smith}
\affiliation{School of Physics and Astronomy, University of Southampton,  Southampton, SO17 1BJ, United Kingdom}
\author{M.~Soares-Santos}
\affiliation{Brandeis University, Physics Department, 415 South Street, Waltham, Massachusetts 02453, USA}
\author{E.~Suchyta}
\affiliation{Computer Science and Mathematics Division, Oak Ridge National Laboratory, Oak Ridge, Tennessee 37831, USA}
\author{M.~E.~C.~Swanson}
\affiliation{National Center for Supercomputing Applications, 1205 West Clark Street, Urbana, Illinois 61801, USA}
\author{G.~Tarle}
\affiliation{Department of Physics, University of Michigan, Ann Arbor, Michigan 48109, USA}
\author{D.~L.~Tucker}
\affiliation{Fermi National Accelerator Laboratory, P. O. Box 500, Batavia, Illinois 60510, USA}
\author{A.~R.~Walker}
\affiliation{Cerro Tololo Inter-American Observatory, NSF's National Optical-Infrared Astronomy Research Laboratory, Casilla 603, La Serena, Chile}
\author{W.~Wester}
\affiliation{Fermi National Accelerator Laboratory, P. O. Box 500, Batavia, Illinois 60510, USA}

\collaboration{DES Collaboration}

\date{\today}

\begin{abstract}
We perform a comprehensive study of Milky Way (MW) satellite galaxies to constrain the fundamental properties of dark matter (DM). 
This analysis fully incorporates inhomogeneities in the spatial distribution and detectability of MW satellites and marginalizes over uncertainties in the mapping between galaxies and DM halos, the properties of the MW system, and the disruption of subhalos by the MW disk. Our results are consistent with the cold, collisionless DM paradigm and yield the strongest cosmological constraints to date on particle models of warm, interacting, and fuzzy dark matter. At $95\%$ confidence, we report limits on (i) the mass of thermal relic warm DM, \CHECK{$m_{\rm WDM} > 6.5\keV$} (free-streaming length, \CHECK{$\lambda_{\rm{fs}} \lesssim 10\,h^{-1}\kpc$}), (ii) the velocity-independent DM-proton scattering cross section, \CHECK{$\sigma_{0} < 8.8\times 10^{-29} \cm^{2}$} for a $100\MeV$ DM particle mass (DM-proton coupling,~\CHECK{$c_p \lesssim (0.3\GeV)^{-2}$}), and (iii) the mass of fuzzy DM, \CHECK{$\mFDM> 2.9 \times 10^{-21}\eV$} (de~Broglie wavelength, \CHECK{$\lambda_{\rm{dB}} \lesssim 0.5\kpc$}). These constraints are complementary to other observational and laboratory constraints on DM properties.
\end{abstract}

\maketitle 

\noindent \emph{Introduction.}---In the concordance model of cosmology, collisionless cold dark matter (CDM) makes up~$\roughly 25\%$ of the matter-energy density of the Universe~\citep{Planck:2019}.
While dark matter (DM) has the potential to solve a number of outstanding challenges in the standard model (SM) of particle physics \citep[][]{Peccei:1977,Jungman:1996, Bertone:2005}, the only positive empirical evidence for DM comes from cosmological and astrophysical observations.
Furthermore, by studying the astrophysical distribution of DM, it is possible to probe its particle nature~\citep{Bullock:2017,Buckley:2018}.
Specifically, the formation, abundance, and structure of gravitationally bound DM structures, known as ``halos,'' provide valuable information about viable ranges of the DM particle mass, production mechanism, and couplings to the SM. 
In particular, the abundance and properties of the smallest DM halos have the potential to indicate a departure from the CDM paradigm~\citep{Bullock:2017,Buckley:2018}.

The smallest known DM halos host the ultrafaint dwarf satellite galaxies of the Milky Way (MW) \citep{Simon190105465}.
In these systems, star formation is highly suppressed by reionization and stellar feedback, leading to mass-to-light ratios that are hundreds of times larger than the universal average \citep{Bullock:2000,Simon190105465}.
Ultrafaint satellite galaxies are, thus, pristine laboratories for studying DM; in particular, the abundance of these systems is a sensitive probe of \emph{any} DM physics that suppresses the formation or present-day abundance of small halos \cite{Maccio:2009,Peter10030419,Polisensky10041459,Anderhalden12122967,Kennedy:2014,Abazajian14030954,Boehm:2014,Safarzadeh190611848}.

Here, we study the following theoretical paradigms for DM that affect the properties of the MW satellite population:
\begin{enumerate}[label=({\roman*})]
    \item \emph{Warm dark matter (WDM)} is produced in the early Universe with a temperature of~$\mathcal{O}(1\keV)$, although its momentum distribution can be nonthermal. Any viable WDM candidate must be cold enough to reproduce the observed large-scale structure, but its non-negligible free-streaming length suppresses the formation of the low-mass halos that host MW satellite galaxies~\citep{Maccio:2009,Polisensky10041459,Anderhalden12122967,Abazajian14030954,Lovell:2014,Kennedy:2014}. One of the most popular WDM candidates is a sterile neutrino~\citep{Abazajian:2017,Adhikari:2017}.

    \item \emph{Interacting dark matter (IDM)} couples strongly enough to the SM to be heated by interactions with the photon-baryon fluid before recombination. This collisional damping washes out small-scale structure, even if the DM is produced nonthermally~\cite{Boehm0012504,Boehm0410591,Nadler190410000}. DM-nucleon interactions arise in generalizations of the weakly-interacting-massive-particle (WIMP) scenario \citep{Boddy:2018,Boddy180800001,Gluscevic1812108}, and the impact of DM-radiation interactions on low-mass halos has also been studied~\citep{Boehm:2014,Schewtschenko151206774,Escudero:2018}. Here, we consider a velocity- and spin-independent DM-proton coupling,~$c_p$.

    \item \emph{Fuzzy dark matter (FDM)} consists of an ultralight boson with a sufficiently small mass, $\mathcal{O}(10^{-22}\eV)$, such that its de~Broglie wavelength is comparable to the sizes of dwarf galaxies, $\mathcal{O}(1\kpc)$; this inhibits the formation of low-mass halos due to the uncertainty principle~\cite{Hu:2000,Hui:2017,Du180104864,Du:2019}. Ultralight axions constitute one popular class of FDM~\cite{Marsh151007633}.
\end{enumerate}

In this Letter, we use novel measurements and modeling of the MW satellite galaxy population to constrain each DM paradigm described above. 
Specifically, we combine a census of MW satellites \cite{PaperI} from the Dark Energy Survey (DES) \cite{DR1:2018} and Pan-STARRS1 (PS1) \cite{Chambers:2016} with a rigorous forward-modeling framework~\cite{PaperII} to fit the position-dependent MW satellite luminosity function in each of these DM paradigms.
This procedure fully incorporates inhomogeneities in the observed MW satellite population and marginalizes over uncertainties in the mapping between MW satellite galaxies and DM halos, the efficiency of subhalo disruption due to the MW disk, and the properties of the MW system.

Our analysis yields stringent constraints on each DM paradigm based on the abundance of observed MW satellites. These limits are complementary to constraints from the Lyman-$\alpha$ forest \citep{Viel:2013,Irsic:2017,Irsic170304683,Rogers200712705}, strongly lensed systems~\citep{Hsueh:2020,Gilman:2020}, and MW stellar streams~\citep{Banik:2019}. 
Our results imply that CDM is consistent with astrophysical observations down to the smallest currently accessible scales ($k \sim 40h \Mpc^{-1}$) and strongly reinforce previous work demonstrating that there is no discrepancy between the number of MW satellites predicted by CDM and current observations \citep{Kim:2018}.
Throughout this work, we fix cosmological parameters at $h = 0.7$, $\Omega_{m} = 0.286$, $\Omega_{\Lambda} = 0.714$, $\sigma_8 = 0.82$, and $n_s=0.96$ \citep{Hinshaw_2013}.

\begin{figure*}[th!]
\includegraphics[width=1.0\textwidth]{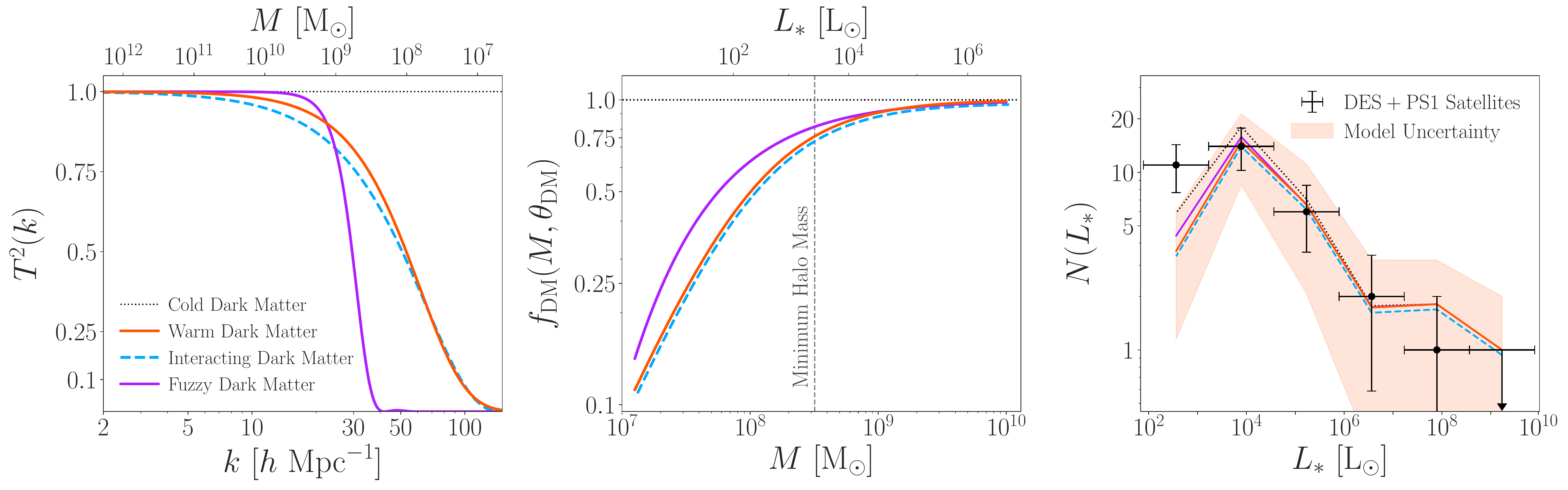}
\caption{\label{fig:sequence} Left panel: Transfer functions for the WDM (orange), IDM (blue), and FDM (magenta) models that are ruled out by our analysis at $95\%$ confidence, corresponding to $\mWDM=6.5\keV$, $\sigma_0=8.8\times 10^{-29}\cm^2$ (for DM particle mass $m_{\chi} = 100\MeV$), and $\mFDM=2.9\times 10^{-21}\eV$, respectively. These constraints are marginalized over our MW satellite model and the properties of the MW system. Middle panel: SHMF suppression relative to CDM for each ruled-out non-CDM model. The vertical dashed line indicates the $95\%$ confidence upper limit on the lowest-mass halo inferred to host MW satellite galaxies \cite{PaperII}. Note that the IDM SHMF is assumed to be identical to the WDM SHMF in our analysis, and is offset slightly for visual clarity. Right panel: Predicted MW satellite galaxy luminosity functions for each ruled-out non-CDM model compared to DES and PS1 observations, evaluated at the best-fit MW satellite model parameters from Ref.\ \cite{PaperII}. The shaded band illustrates the uncertainty of our WDM prediction due to the stochasticity of our galaxy-halo connection model and the limited number of simulations used in our analysis; the size of this uncertainty is very similar to that in CDM and the other alternative DM models shown. This panel is a simple one-dimensional representation of our MW satellite and DM model fit to the luminosity, size, and spatial distribution of satellites in the DES and PS1 survey footprints. The comparison of our CDM model to data is described in Ref.\ \cite{PaperII}, and full posterior distributions for our non-CDM analyses are provided in Supplemental Material.}
\end{figure*}

\vspace{1em} \noindent \emph{Analysis overview.}---
Before discussing our treatment of each DM paradigm in detail, we describe the main components of our analysis used to connect non-CDM scenarios to the observed MW satellite population. For each paradigm, we assume that the non-CDM component constitutes the entirety of the DM. Figure \ref{fig:sequence} illustrates how our analysis proceeds: Non-CDM physics suppresses the linear matter power spectrum on small scales (left panel), which manifests as an underabundance of subhalos (middle panel) and faint MW satellite galaxies (right panel) relative to CDM predictions.

\emph{Transfer function.}---The linear matter power spectrum, normalized to that of CDM, is used to generate initial conditions for simulations of structure formation. In particular, the transfer function is defined as
    \begin{equation}
        T^2(k)\equiv \frac{P_\DM(k)}{P_\CDM(k)},
    \end{equation}
    where $k$ is the cosmological wave number, $P_{\rm CDM}(k)$ is the CDM linear matter power spectrum, and $P_{\rm DM}(k)$ is the linear matter power spectrum of a non-CDM model~\citep[][]{Bode:2000gq}.~$P_{\rm DM}(k)$ is obtained by integrating the relevant Boltzmann equation (which may include DM-SM interactions) given the initial DM phase-space distribution. The left panel in \figref{sequence} illustrates the transfer function for the three DM paradigms we consider.
    
    It is convenient to define the \emph{half-mode scale} \khm as the wave number satisfying $T^2(\khm)=0.25$ \cite{Schneider11120330}. The corresponding \emph{half-mode mass},
    \begin{equation}
    \Mhm = \frac{4\pi}{3}\Omega_{m}\bar{\rho}\left(\frac{\pi}{\khm}\right)^3
    \end{equation}
    is a characteristic mass scale below which the abundance of DM halos is significantly suppressed relative to CDM. Here, $\bar{\rho}$ is the critical density of the Universe today.

\emph{Subhalo mass function (SHMF).}---The abundance of subhalos within the virial radius of the MW is expressed as the cumulative number of subhalos as a function of subhalo mass $M$.
    We follow Ref.\ \cite{Nadler190410000} by using peak virial mass, defined according to the Bryan-Norman overdensity \cite{Bryan_1998} with $\Delta_{\rm vir}\simeq 99.2$ (consistent with our cosmological parameters). 
    We define
    \begin{equation}
        \left(\frac{dN_{\mathrm{sub}}}{dM}\right)_\DM \equiv f_\DM(M,\vect{\theta}_\DM)\left(\frac{dN_{\mathrm{sub}}}{dM}\right)_\CDM,\label{eq:shmf}
    \end{equation}
    where $f_\DM(M,\vect{\theta}_\DM)$ is the suppression of the SHMF relative to CDM and $\vect{\theta}_\DM$ are DM model parameters; both $f_\DM$ and $\vect{\theta}_\DM$ depend on the DM model in question. The middle panel in \figref{sequence} shows SHMF suppression for the three DM paradigms we consider.

\emph{MW satellite model.}---Here, we describe the additions to our MW satellite model pertaining to the non-CDM paradigms described above. We comprehensively discuss the underlying galaxy-halo connection model in Supplemental Material. We combine the SHMF suppression in \eqnref{shmf} with a state-of-the-art satellite modeling framework \cite{PaperII} to predict the abundance of observed MW satellites in each DM paradigm. Our modeling framework combines cosmological zoom-in simulations of two halos from Ref.\ \cite{Mao150302637}---which are
chosen to have masses, concentrations, and assembly histories similar to those inferred for the MW halo and include realistic analogs of the Large Magellanic Cloud system---with a statistical model of the galaxy-halo connection in order to populate subhalos with satellite galaxies.
    
We implement SHMF suppression by multiplying the detection probability of each mock satellite, which includes terms that model tidal disruption due to the MW disk, the efficiency of galaxy formation, and observational detectability, by a factor of $f_\DM(M,\vect{\theta}_\DM)$, following Refs.\ \citep{Jethwa161207834,Nadler190410000}. This procedure assumes that the shape of the observed radial satellite distribution (which our model predicts reasonably well \cite{PaperII}) is unchanged in alternative DM scenarios, which is consistent with results from cosmological WDM simulations of MW-mass halos~\cite{Lovell:2014,Bose160407409}. The validity of this assumption is less certain for FDM, because dynamical friction operates differently for wavelike versus particle DM \cite{Lancaster190906381}, although this results in negligible differences in disruption timescales for the $\sim 10^8 \Msun$ subhalos that drive our constraints~\citep{Du180104864}. The right panel in \figref{sequence} shows the predicted satellite luminosity function for each non-CDM model under consideration evaluated with model parameters that are ruled out at $95\%$ confidence.
    
\emph{Fitting procedure.}---We fit predicted satellite populations to the observed satellite population from DES and PS1 using the observational selection functions derived in Ref.\ \cite{PaperI}, assuming that satellite surface brightness is distributed according to a Poisson point process in each survey footprint \citep{Nadler:2018,PaperII}. We use the Markov chain Monte Carlo (MCMC) code \texttt{EMCEE}~\citep{emcee} to simultaneously fit for seven parameters governing the galaxy-halo connection, one parameter governing the impact of the MW disk on subhalo disruption, and one parameter governing the impact of the DM model in question, which we express as a subhalo mass scale. In particular, our thermal relic WDM constraint is derived by fitting for $\Mhm$, and our FDM limit is derived by fitting for a characteristic mass scale $M_0$. Further details on our fitting procedure are provided in Supplemental Material.
    
    Subhalo abundance is known to scale linearly with host halo mass \cite{Mao150302637}, and we assume that satellite luminosity is a monotonic function of subhalo mass, modulo scatter \cite{PaperII}. We therefore expect a higher-mass MW host halo to yield weaker constraints on non-CDM models, because observed satellites would inhabit correspondingly higher-mass subhalos. The average virial mass of the host halos in our two realistic MW-like simulations is $1.4\times 10^{12}\Msun$, which is consistent with the $95\%$ confidence range for the virial mass of the MW halo inferred from \emph{Gaia} measurements of satellite kinematics \citep{Callingham180810456,Cautun191104557}. To be conservative, we account for the uncertainty in MW halo mass on our DM constraints by assuming that the mass scale describing the suppression of the SHMF in each DM paradigm is linearly related to the virial mass of the MW halo, following the scaling for minimum halo mass derived in Ref.\ \cite{PaperII}. In particular, we multiply the upper limit on the characteristic mass scale in each of our non-CDM fits by the ratio of the largest allowed MW halo mass to the average host halo mass in our simulations. We validate this procedure by fitting the observed satellite population using each of our two MW-like simulations separately, which yields reasonable agreement with the linear scaling expectation. This conservative scaling mitigates the largest uncertainty associated with the limited statistics of our two realistic simulations. 
    
    In summary, our fit to the MW satellite population incorporates both intrinsic inhomogeneities in the spatial distribution of MW satellites and those introduced by the varying coverage and depth of current surveys. We assume that alternative DM physics modifies only the SHMF, via \eqnref{shmf}, and we report $95\%$ confidence limits on DM model parameters that are marginalized over uncertainties in our MW satellite model and the properties of the MW system.
    
\begin{table*}[t]
\begin{center}
\begin{tabular*}{\textwidth}{l l c l c}
\hline
Dark matter paradigm & Parameter & Constraint & Derived property & Constraint\\
\hline
\hline\\[-0.5em]
Warm dark matter & Thermal relic mass & \CHECK{$\mWDM > 6.5\keV$} & Free-streaming length & \CHECK{$\lambda_{\rm{fs}} \lesssim 10\,h^{-1}\kpc$}  \\
& & & &\\
Interacting dark matter~~ & \vtop{\hbox{\strut Velocity-independent}\hbox{\strut DM-Proton cross section~~}} & \CHECK{$\sigma_0 < 8.8\times 10^{-29} \cm^2$}~~ & DM-Proton coupling~~ & \CHECK{$c_p \lesssim (0.3\GeV)^{-2}$} \\
Fuzzy dark matter & Particle mass & \CHECK{$\mFDM > 2.9 \times 10^{-21} \eV$} & de Broglie wavelength & \CHECK{$\lambda_{\rm{dB}} \lesssim 0.5\kpc$} \\[+0.5em]
\hline
\end{tabular*}
\end{center}
\caption{\label{tab:limits}
Constraints on the WDM, IDM, and FDM paradigms from observations of MW satellite galaxies. 
Limits for each non-CDM model are derived by assuming that it constitutes the entirety of the DM. 
The first column lists the DM paradigm, the second column describes the particle physics parameters constrained by this analysis, the third column lists the corresponding constraints at 95\% confidence, the fourth column describes the derived property constrained for each DM model, and the fifth column lists constraints on the derived parameters. 
Limits on the DM-proton scattering cross sections depend on the DM particle mass, $m_{\chi}$ (see \figref{limits}); for simplicity, we present our constraint for $m_{\chi} = 100\MeV$.
}
\end{table*}

\emph{WDM Analysis.}---Thermal relic WDM with particle mass, \mWDM, has been studied extensively in the literature (e.g., Refs.\ \cite{Viel0501562,Lovell:2014}) and serves as a benchmark model for our analysis.

\emph{Transfer function.}---The transfer function for thermal relic WDM is given as a function of \mWDM by Ref.\ \cite{Viel0501562}.
This transfer function is commonly assumed in cosmological studies of WDM and facilitates a well-defined comparison to other small-scale structure results~\cite{Viel:2013,Irsic:2017,Hsueh:2020,Gilman:2020,Banik:2019}.
However, the simple thermal relic transfer function is inadequate to describe specific particle models of WDM, such as resonantly produced sterile neutrinos~\cite{Lovell191111785}.
Thus, constraints on specific DM candidates must be inferred using transfer functions appropriate for the particle model in question, as we discuss below.

\emph{SHMF.}---Several authors have implemented the thermal relic WDM transfer function from Ref.\ \cite{Viel0501562} in cosmological zoom-in simulations to estimate the suppression of the SHMF in MW-mass host halos \citep{Schneider11120330,Angulo13042406,Lovell:2014,Bose160407409}. 
These results depend on the algorithm used to remove spurious halos \citep{Wang0702575,Angulo13042406} and, therefore, vary among studies. 
Following Ref.\ \cite{Lovell200301125}, SHMF suppression for thermal relic WDM can be expressed as
    \begin{equation}
    f_\WDM(M,\mWDM) = \left[1+\left(\frac{\alpha\Mhm(\mWDM)}{M}\right)^{\beta}\right]^{\gamma},\label{eq:wdm_shmf}
\end{equation}
where $\alpha$, $\beta$, and $\gamma$ are constants and $\Mhm$ is related to $\mWDM$ in our fiducial cosmology via
\begin{equation}
    \Mhm(\mWDM) = 5 \times 10^8 \left(\frac{\mWDM}{3 \keV}\right)^{-10/3} \Msun.\label{eq:mhm_mwdm}
\end{equation}

To facilitate comparison with recent WDM constraints from analyses of the MW satellite population \cite{Nadler190410000}, strong gravitational lenses \cite{Hsueh:2020,Gilman:2020}, and stellar streams \cite{Banik:2019}, we adopt the SHMF from Ref.\ \cite{Lovell:2014}, which corresponds to \eqnref{wdm_shmf} with $\alpha=2.7$, $\beta=1.0$, and $\gamma=-0.99$. We note that the recent estimate of the SHMF from Ref.\ \cite{Lovell200301125}---which specifically models resonantly produced sterile neutrino WDM---is significantly \emph{less} suppressed than the thermal relic SHMF from Ref.\ \cite{Lovell:2014}. Thus, our fiducial WDM constraint applies directly only to thermal relic DM.

\emph{Fitting procedure.}---We implement \eqnref{wdm_shmf} in our fit to the MW satellite population to obtain a marginalized posterior distribution over~$\Mhm$. In particular, we fit for $\log_{10}(\Mhm)$ using a uniform prior on this logarithmic quantity, and we translate the resulting limit to \mWDM using \eqnref{mhm_mwdm}. We translate our thermal relic WDM limit into constraints on resonantly produced sterile neutrinos by following Refs.\ \cite{Schneider160107553,Maamari}. Specifically, we analyze sterile neutrino transfer functions over a grid of mass and mixing angle values \cite{Cherry:2017}, and we constrain sterile neutrino models that produce transfer functions which are strictly more suppressed than our $95\%$ confidence ruled-out thermal relic WDM model. This procedure is described in detail in Supplemental Material.

\emph{IDM Analysis.}---Our treatment of IDM follows the prescription of Ref.\ \cite{Nadler190410000}. For concreteness, we focus on the case of velocity-independent DM-proton scattering.

\emph{Transfer function.}---Following Ref.\ \cite{Nadler190410000}, the transfer function in our fiducial IDM model is obtained using the modified version of the Boltzmann solver CLASS described in Refs.\ \cite{Boddy:2018,Boddy180800001,Gluscevic1812108}, which we use to evolve linear cosmological perturbations in the presence of velocity-independent DM-proton interactions. These interactions are described by the velocity-independent scattering cross section $\sigma_0$ and the DM particle mass $m_{\chi}$. As noted in Ref.\ \cite{Nadler190410000}, transfer functions for this model are very similar to those of thermal relic WDM, modulo dark acoustic oscillations that occur at very small scales and are significantly suppressed for our parameter space of interest.

\emph{SHMF.}---Because cosmological zoom-in simulations including DM-proton scattering have not been performed, we follow Ref.\ \cite{Nadler190410000} by mapping the SHMF suppression of IDM to that of WDM based on the correspondence of the transfer functions. In particular, we match the half-mode scales in the transfer functions to construct a relation between \mWDM and $(\sigma_0,m_{\chi})$, and we assume that the IDM SHMF is identical to the corresponding thermal relic WDM SHMF from Ref.\ \cite{Lovell:2014}. 
This procedure neglects late-time DM-proton scattering, which has a negligible impact on subhalo abundances in our IDM model, even in regions with high baryon densities.

\emph{Fitting procedure.}---Following Ref.\ \cite{Nadler190410000}, we use the mapping procedure described above to translate our $95\%$ confidence limit on thermal relic WDM into limits on $\sigma_0$ for several values of $m_\chi$ in our fiducial IDM model.

\emph{FDM Analysis.}---Finally, we provide details on each step for the FDM paradigm. We focus on the case of ultralight scalar field DM with negligible self-interactions and SM couplings.

\emph{Transfer function.}---The FDM transfer function is given as a function of the FDM mass $\mFDM$ by Ref.\ \cite{Hu:2000}. 
We note that this transfer function features steeper power suppression than thermal relic WDM for a fixed half-mode scale.

\emph{SHMF.}---We assume that the FDM SHMF suppression takes the form of \eqnref{shmf}, and we fit the results of the semianalytic model in Refs.\ \cite{Du180104864,Du:2019} with a function of the form
\begin{equation}
    f_\FDM(M,m_\phi) = \left[1+\left(\frac{M_{0}(\mFDM)}{M}\right)^{\tilde{\beta}(\mFDM)}\right]^{\tilde{\gamma}(\mFDM)},\label{eq:fdm_shmf}
\end{equation}
where $\tilde{\beta}(\mFDM)$ and $\tilde{\gamma}(\mFDM)$ are provided in Supplemental Material.
The characteristic subhalo mass scale $M_0$ is related to the FDM mass via \cite{Schive150804621}
\begin{equation}
    M_0(\mFDM) = 1.6\times 10^{10} \left(\frac{\mFDM}{10^{-22}\eV}\right)^{-4/3}\Msun.\label{eq:m0_fdm}
\end{equation}

The SHMF suppression in \eqnref{fdm_shmf} encapsulates the effects of tidal stripping on subhalos with solitonic cores, which was explicitly included by Refs.\ \cite{Du180104864,Du:2019}. 
This SHMF suppression is significantly less severe than that estimated from the FDM simulations in Ref.\ \cite{Schive150804621}. As described in Supplemental Material, using the SHMF from Ref.\ \cite{Schive150804621} in our fit yields a limit on the FDM mass that is roughly 3 times more stringent than our fiducial result. 
This confirms that the FDM SHMF is a key theoretical uncertainty that must be addressed \cite{Hui:2017}.

\emph{Fitting procedure.}---We implement the SHMF in \eqnref{fdm_shmf} in our fit to the MW satellite population to obtain a marginalized posterior distribution over $M_{0}$. In particular, we fit for $\log_{10}(M_{0})$ using a uniform prior on this logarithmic quantity, and we translate the resulting limit to $\mFDM$ using \eqnref{m0_fdm}. We note that our procedure for constraining FDM uses the detailed \emph{shape} of the SHMF suppression in this model rather than mapping the half-mode scale of the FDM transfer function to that of thermal relic WDM as in Ref.\ \cite{Nadler190410000} or bounding the FDM SHMF by ruled-out thermal relic WDM SHMFs as in Ref.\ \cite{Schutz200105503}. This is necessary because both the shape of the FDM transfer function and the resulting suppression of the SHMF differ in detail from thermal relic WDM (see \figref{sequence}).

\begin{figure*}[t]
\includegraphics[width=0.495\textwidth]{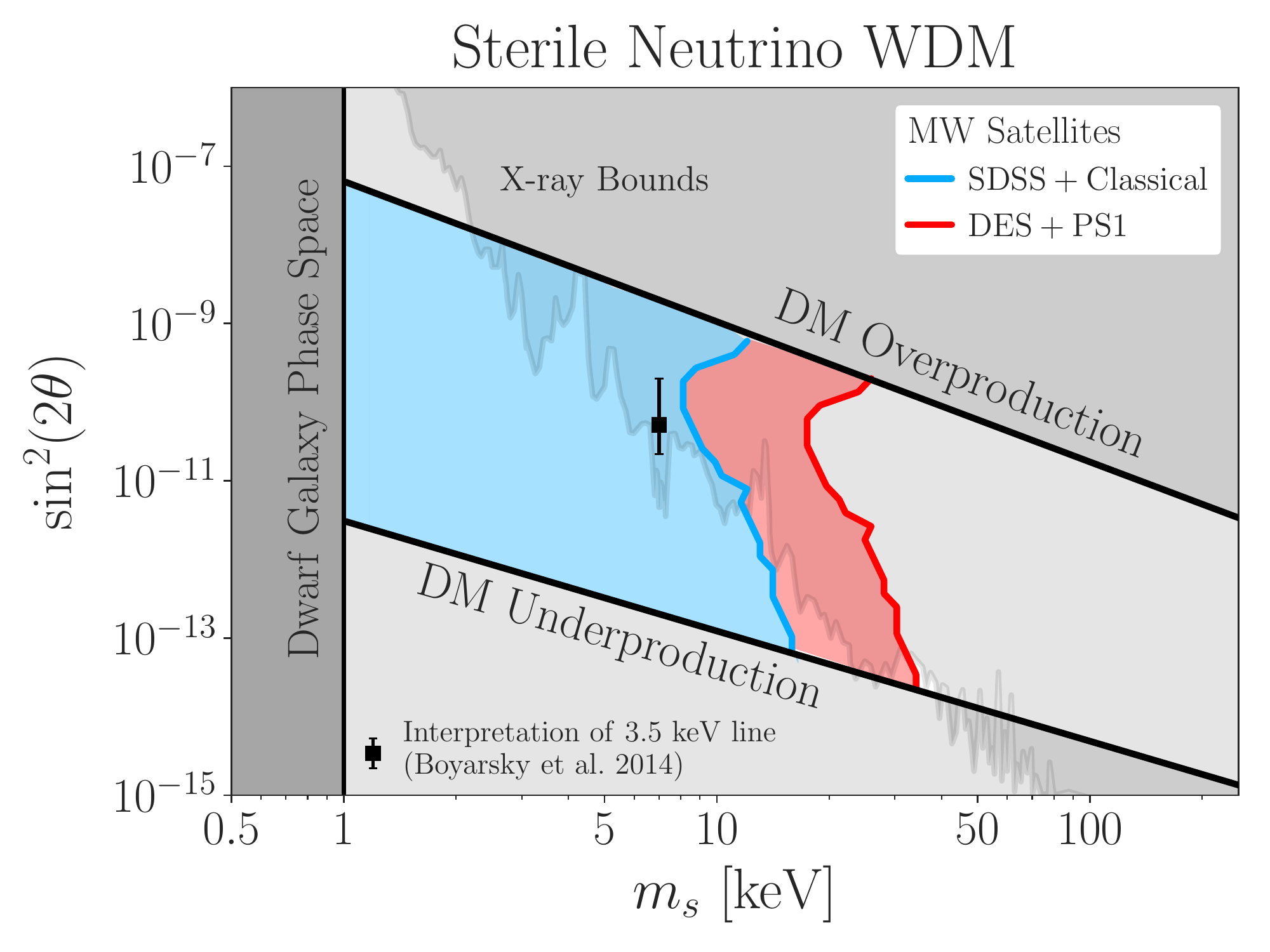}
\includegraphics[width=0.495\textwidth]{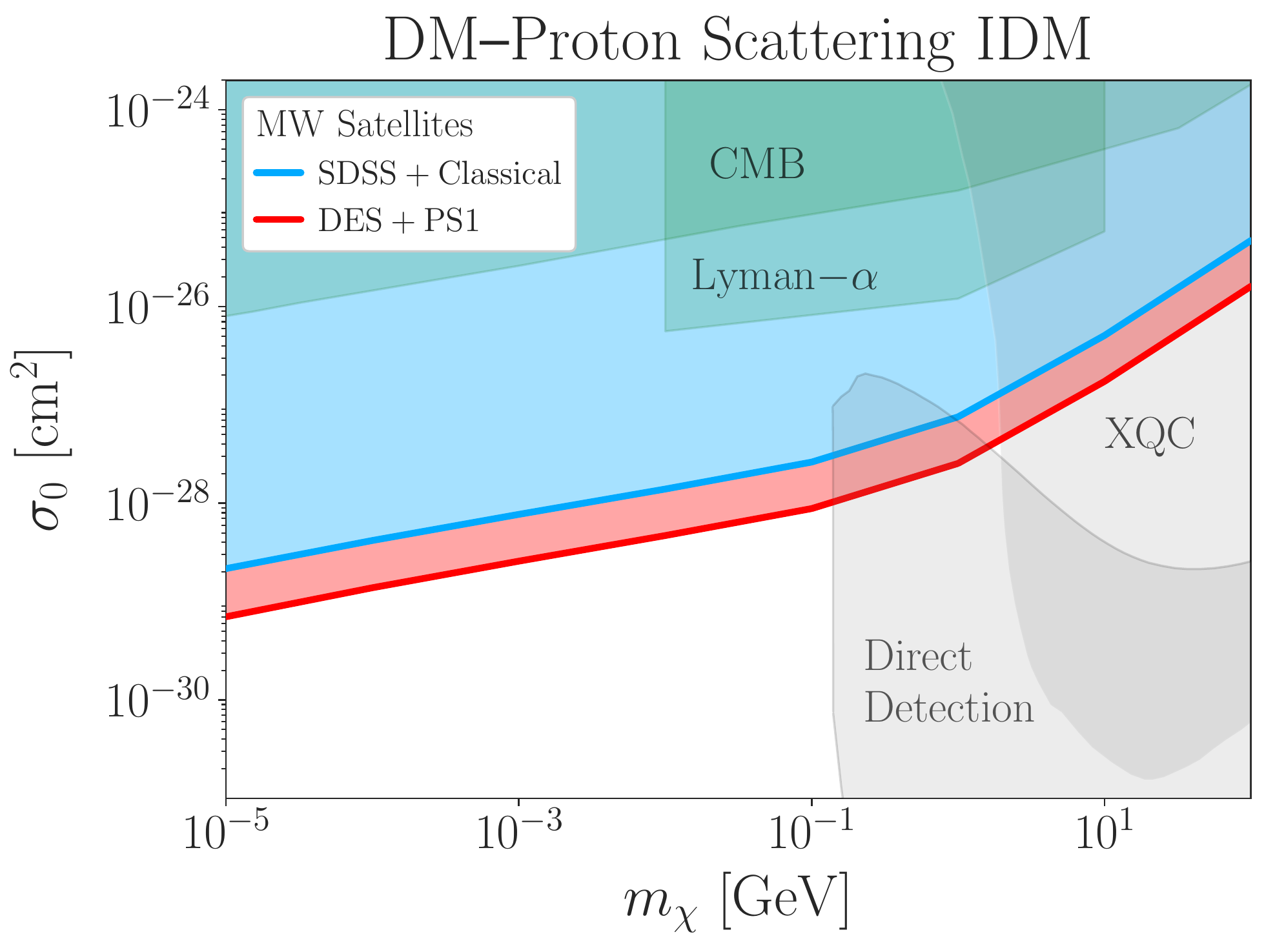}
\caption{Exclusion regions for WDM and IDM models from our analysis of MW satellites observed with DES and PS1 (red) compared to previous constraints from classical and SDSS satellites \citep{Nadler190410000} (blue) and other experimental results. Left panel: Constraints on the mass and mixing angle of resonantly produced sterile neutrino DM. These constraints are derived by finding mass and mixing angle combinations that suppress the linear matter power spectrum more strongly than the $\mWDM=6.5\keV$ thermal relic ruled out at $95\%$ confidence by our analysis. The black point with error bars shows the sterile neutrino interpretation of the 3.5\keV x-ray line \cite{Boyarsky:2014}. The dark gray region is ruled out by dwarf galaxy internal dynamics \cite{Boyarsky08083902}, and the gray contour shows x-ray constraints \citep{Horiuchi:2014,Perez:2017,Dessert1465}. 
Solid black lines indicate regions of parameter space in which resonantly produced sterile neutrinos cannot constitute all of the DM in the neutrino minimal standard model \citep{Asaka:2005,Schneider160107553}. 
Right panel: Constraints on the interaction cross section and DM mass for velocity-independent DM-proton scattering. Green contours show cosmological limits from the CMB \citep{Boddy:2018,Gluscevic1812108} and the Lyman-$\alpha$ forest \citep{Xu:2018}. 
Light gray contours show experimental limits from the x-ray quantum calorimeter~\citep{Mahdawi180403073} and direct detection results 
as interpreted by Ref.\ \cite{Emken:2018}.\label{fig:limits}}
\end{figure*}

\begin{figure}[t]
\hspace{-6.5mm}
\includegraphics[width=0.48\textwidth]{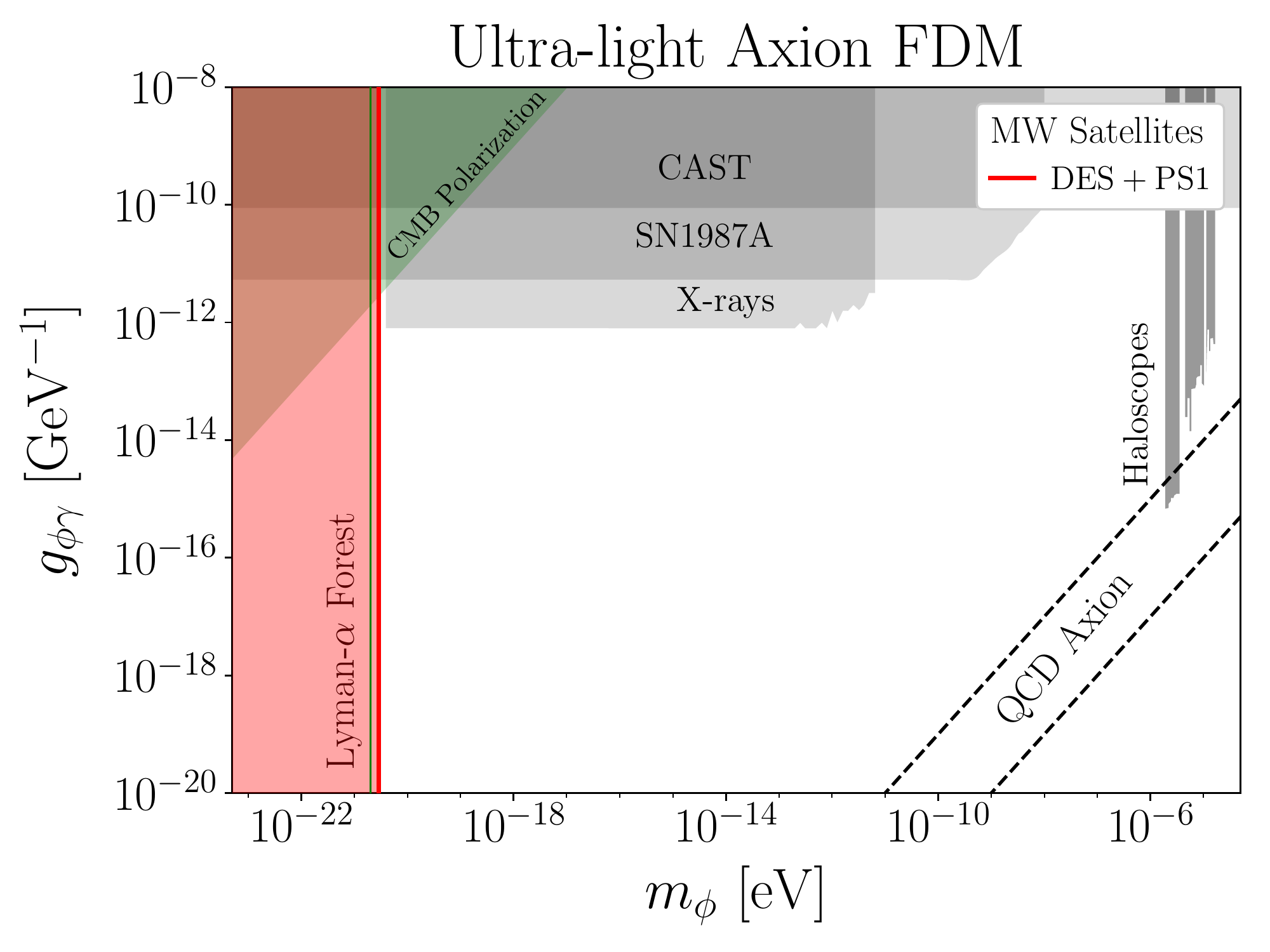}
\caption{Constraints on ultralight axion particle mass versus axion-photon coupling from our analysis of the MW satellite population (red).
Limits from CMB polarization washout \cite{Fedderke:2019} and the Lyman-$\alpha$ forest \citep{Irsic170304683} are shown in green, and haloscope limits are shown as gray vertical bands.
Experimental constraints from the CAST experiment \citep{Arik:2015}, the lack of a $\gamma$-ray signal from  SN1987A \cite{Payez:2015}, and the x-ray transparency of the intracluster medium \citep{Reynolds:2019} are shown in gray and do not require that the ultralight axion makes up all of the DM. The dashed lines \cite{Ringwald13101256} span canonical QCD axion models \citep{Kim:1979,Dine:1981}.
\label{fig:fdm_limits}}
\end{figure}

\vspace{1em} \noindent \emph{Results.}---\tabref{limits} presents our constraints on the WDM, IDM, and FDM paradigms. We describe these results below and translate the limits into constraints on specific models corresponding to each DM paradigm.
\begin{enumerate}[label=({\roman*})]
\item \emph{WDM}.---Our fit using the thermal relic WDM SHMF suppression from Ref.\ \cite{Lovell:2014} yields \CHECK{$\Mhm < 3.0\times 10^7\Msun$}, or \CHECK{$\mWDM>7.0\keV$}, at $95\%$ confidence. Linear scaling with MW halo mass yields our fiducial constraint of~\CHECK{$\Mhm < 3.8\times 10^7\Msun$}, corresponding to \CHECK{$\mWDM>6.5\keV$}. This translates to an upper limit on the free-streaming length of~\CHECK{$\lambda_{\rm{fs}} \lesssim 10\ h^{-1}\kpc$}, corresponding to the virial radii of the smallest halos that host MW satellite galaxies, and improves on previous $\mWDM$ constraints from the MW satellite population by a factor of $\roughly 2$~\cite{Nadler190410000}.

Our constraint on thermal relic WDM translates to a lower limit of $50\keV$ on the mass of a nonresonant Dodelson-Widrow sterile neutrino \cite{Dodelson:1994, Viel0501562}.
We also translate our thermal relic WDM limit into constraints on the mass and mixing angle of resonantly produced sterile neutrinos assuming a Shi-Fuller production mechanism \cite{Shi:1999}, following the conservative procedure described above. 
As shown by the red exclusion region in the left panel in \figref{limits}, our analysis rules out nearly the entire remaining parameter space for resonantly produced sterile neutrinos in the neutrino minimal standard model \citep{Asaka:2005} at greater than $95\%$ confidence. (A small region of parameter space is not excluded at the lowest viable mixing angles and $m_s\gtrsim 30\ \mathrm{keV}$.) 
In addition, we robustly rule out the resonantly produced sterile neutrino interpretation of the $3.5\keV$ x-ray line~\cite{Boyarsky:2014}.

\item \emph{IDM}.---Mapping our \CHECK{$\mWDM>6.5\keV$} constraint to the DM-proton scattering model following the procedure in Ref.\ \cite{Nadler190410000} yields constraints on the velocity-independent interaction cross section of \CHECK{$(7.0\times 10^{-30},2.6\times 10^{-29},8.8\times 10^{-29}, 1.7\times 10^{-27})\cm^2$} for DM particle masses of~\CHECK{$(10^{-5}, 10^{-3}, 10^{-1}, 10)\GeV$}, respectively, at $95\%$ confidence. As shown by the red exclusion region in the right panel in \figref{limits}, these constraints are highly complementary to direct detection limits, particularly at low DM masses~\cite{Nadler190410000}. We note that these constraints scale as $m_{\chi}^{1/4}$ ($m_{\chi}$) for $m_{\chi}\ll 1\GeV$ ($m_{\chi}\gg 1\GeV$). At a DM mass of $100\MeV$, our limit translates into an upper bound on the DM-proton coupling of \CHECK{$c_p \lesssim (0.3\GeV)^{-2}$} \cite{Boddy:2018}.

Despite our conservative marginalization over MW halo mass, these results improve upon those in Ref.\ \cite{Nadler190410000} by a factor of \CHECK{$\sim 3$} at all DM masses. This is stronger than the improvement expected from the analytic prediction for cross section constraints derived in Ref.\ \cite{Nadler190410000} due to a more precise determination of the SHMF, resulting from the sky coverage and sensitivity of DES and PS1.

Several complementary astrophysical and cosmological measurements probe the DM-proton scattering cross section. Stringent limits have been derived by reinterpreting direct detection constraints in the context of cosmic ray upscattering \citep{Bringmann:2019}. We do not show these results in \figref{limits}, because they constrain the DM-proton scattering at relativistic energies, which precludes a straightforward mapping to the velocity-independent cross section constrained here. The IDM model we consider contributes to the energy density of relativistic species at big bang nucleosynthesis, which sets a lower on its mass that depends on the spin statistics of the DM particle \cite{Boehm13036270,Nollett14116005,Krnjaic190800007}. Understanding the interplay of these results with our limits is an important area for future work.

\item \emph{FDM}.---We obtain \CHECK{$M_{0}< 1.4 \times 10^8\Msun$} at $95\%$ confidence from our fiducial FDM fit.  
Applying linear MW-host mass scaling yields \CHECK{$M_{0}<1.8\times 10^8\Msun$} at~$95\%$ confidence, or \CHECK{$\mFDM>2.9\times 10^{-21}\eV$}. This translates to an upper limit on the de Broglie wavelength of~\CHECK{$\lambda_{\rm{dB}} \lesssim 0.5\ h^{-1}\kpc$}, roughly corresponding to the sizes of the smallest MW satellite galaxies. Thus, the $10^{-22}\eV$ FDM model invoked to reconcile the apparent mismatch between the predicted and observed inner dark matter density profiles of dwarf galaxies \cite{Hui:2017}, and to fit the internal dynamics of low-surface-brightness \cite{Bernal:2017,Broadhurst:2020} and ultradiffuse \cite{Wasserman:2019} galaxies, is strongly disfavored by MW satellite abundances.

To connect to particle models of FDM, we plot this limit in the well-motivated parameter space of ultralight axion mass versus axion-photon coupling in \figref{fdm_limits}. For the range of axion-photon couplings that we consider, this mixing has a negligible effect on structure formation. We reiterate that our constraint was derived assuming a light scalar field without self-interactions; this assumption may be violated in specific ultralight axion models. Although our analysis and Lyman-$\alpha$ forest studies exclude a similar region of parameter space \cite{Irsic170304683,Rogers200712705}, our work probes structure on complementary physical scales with distinct theoretical and observational systematics.
\end{enumerate}

\noindent \emph{Discussion.}---In this Letter, we used a state-of-the-art model of the MW satellite galaxy population to place stringent and robust limits on three fundamental DM paradigms: WDM, IDM, and FDM. Although some of these alternative DM models gained popularity by solving apparent small-scale structure ``challenges'' facing CDM, recent observational and theoretical advances have reversed this scenario. 
In particular, astrophysical and cosmological observations of the smallest DM structures are now among the strongest constraints on the microphysical properties of DM.

This analysis improves upon previous work by using MW satellite observations over nearly the entire sky and rigorously accounting for both satellite detectability and uncertainties in the galaxy-halo connection. Our constraints are comparable in sensitivity to Lyman-$\alpha$ forest, strong lensing, and stellar stream perturbation analyses. Future cosmic surveys promise to further improve these measurements and to enable a detailed comparisons to the internal dynamics of these galaxies \citep{Drlica-Wagner:2019,MSE:2019}.

As the observational data improve, there are several uncertainties in the current modeling framework that are important to address. In particular, our use of only two realistic MW simulations limits the range of host halos and subhalo populations that enter our analysis; sampling a wider range of host halo masses, formation histories, and environments will improve the accuracy and precision of DM constraints derived from MW satellite galaxies. We describe other model uncertainties in Supplemental Material.

The breadth of DM models constrained by observations of MW satellites is particularly important given the growing interest in a wide range of theoretical possibilities following nondetections in collider, direct, and indirect searches for canonical WIMPs. In addition to the three DM paradigms considered in this work, small-scale structure measurements are also sensitive to the initial DM velocity distribution in nonthermal production scenarios \citep{Miller190810369}, the DM formation epoch \cite{Sarkar14107129,Das201001137}, the DM self-interaction cross section \citep{Vogelsberger12015892,Zavala12116426,Tulin170502358,Nadler200108754}, and the DM particle lifetime \cite{Peter10030419,Wang14060527}.

Future work could generalize our approach by measuring deviations in the small-scale linear matter power spectrum relative to a baseline CDM scenario rather than setting constraints in the context of particular DM models. 
Features in the power spectrum on extremely small scales are a hallmark of many inflationary models~\cite{Kamionkowski:2000,White:2000}, and it is conceivable that DM substructure measurements can be used to infer the nature of the corresponding primordial density fluctuations.

Our code and subhalo catalogs are available online \cite{PaperIIIcode}.

This Letter has gone through internal review by the DES Collaboration. 
We thank J.~J.\ Cherry and Aurel Schneider for providing transfer functions for resonantly produced sterile neutrinos and Peter Graham for helpful discussions on ultralight axions. This work was supported in part by U.S. Department of Energy contracts to SLAC (DE-AC02-76SF00515) and Fermilab (DE-AC02-07CH11359). Support was received from the National Science Foundation (NSF) under Grant No.\ NSF AST-1517422, Grant No.\ NSF PHY17-48958 through the Kavli Institute for Theoretical Physics program ``The Small-Scale Structure of Cold(?)\ Dark Matter,'' and Grant No.\ NSF DGE-1656518 through the NSF Graduate Research Fellowship received by E.~O.~N. 
Support for Y.-Y.~M.\ and T.~S.~L.\ was provided by NASA through the NASA Hubble Fellowship Grant No.\ HST-HF2-51441.001 awarded by the Space Telescope Science Institute, which is operated by the Association of Universities for Research in Astronomy, Incorporated, under NASA Contract No.\ NAS5-26555.

This research made use of computational resources at SLAC National Accelerator Laboratory, a U.S.\ Department of Energy Office of Science laboratory; the authors are thankful for the support of the SLAC computational team.
This research made use of the Sherlock
cluster at the Stanford Research Computing Center (SRCC); the authors are thankful for the support of the SRCC team. 
This research made use of arXiv.org and NASA's Astrophysics Data System for bibliographic information.

Funding for the DES Projects has been provided by the DOE and NSF (USA), MEC/MICINN/MINECO (Spain), STFC (United Kingdom), HEFCE (United Kingdom). NCSA (UIUC), KICP (University of Chicago), CCAPP (Ohio State), 
MIFPA (Texas A\&M), CNPQ, FAPERJ, FINEP (Brazil), DFG (Germany), and the collaborating institutions in DES.

The collaborating institutions are Argonne Lab, UC Santa Cruz, University of Cambridge, CIEMAT-Madrid, University of Chicago, University College London, 
DES-Brazil Consortium, University of Edinburgh, ETH Z{\"u}rich, Fermilab, University of Illinois, ICE (IEEC-CSIC), IFAE Barcelona, Lawrence Berkeley Lab, 
LMU M{\"u}nchen and the associated Excellence Cluster Universe, University of Michigan, NFS's NOIRLab, University of Nottingham, The Ohio State University, University of Pennsylvania, University of Portsmouth, SLAC, Stanford University, University of Sussex, Texas A\&M University, and the OzDES Membership Consortium.

Based in part on observations at Cerro Tololo Inter-American Observatory at NSF’s NOIRLab (NOIRLab Prop. ID 2012B-0001; PI: J. Frieman), which is managed by the Association of Universities for Research in Astronomy (AURA) under a cooperative agreement with the NSF.

The DES Data Management System is supported by the NSF under Grants No.\ AST-1138766 and No.\ AST-1536171. 
The DES participants from Spanish institutions are partially supported by MICINN under Grants No.\ ESP2017-89838, No.\ PGC2018-094773, No.\ PGC2018-102021, No.\ SEV-2016-0588, No.\ SEV-2016-0597, and No.\ MDM-2015-0509, some of which include ERDF funds from the European Union. IFAE is partially funded by the CERCA program of the Generalitat de Catalunya.
Research leading to these results has received funding from the European Research
Council under the European Union's Seventh Framework Program (FP7/2007-2013) including ERC Grant Agreements No.\ 240672, No.\ 291329, and No.\ 306478.
We  acknowledge support from the Brazilian Instituto Nacional de Ci\^encia
e Tecnologia (INCT) e-Universe (CNPq Grant No.\ 465376/2014-2).

This manuscript has been authored by Fermi Research Alliance, LLC under Contract No. DE-AC02-07CH11359 with the U.S. Department of Energy, Office of Science, Office of High Energy Physics.

\bibliography{main_copy}

\clearpage
\onecolumngrid
\appendix

\section{SUPPLEMENTAL MATERIAL}
\label{sec:supplement}

\subsection{Milky Way Satellite Model}

Here, we provide a high-level overview of our MW satellite galaxy model and its associated uncertainties; we refer the reader to \cite{Nadler:2018,PaperII} for a complete description. Our model of the MW satellite population is based on dark matter-only zoom-in simulations chosen to resemble the MW halo in terms of its mass, concentration, and formation history (namely, a major merger with mass ratio and infall time similar to observational estimates for the \emph{Gaia}-Enceladus merger, and a quiescent mass accretion history thereafter). The MW host halos in these simulations respectively have virial masses of $1.57\times 10^{12}\Msun$ and $1.26\times 10^{12}\Msun$ and concentration values of $11.8$ and $10.5$. These concentration values are consistent within $\sim 1\sigma$ with the range derived in \cite{Callingham180810456}, and they are also consistent with the range inferred by \cite{Cautun191104557} for the concentration of the MW halo before adiabatic contraction in the presence of baryons. This implies that our MW-like host halos are less concentrated than the real MW halo, which potentially impacts the efficiency of subhalo disruption beyond the leading-order effects captured by our model of subhalo disruption due to the MW disk, which is discussed below.

Crucially, each of these two simulations includes a realistic Large Magellanic Cloud analog system (i.e., a system with a total mass, infall time, and orbit consistent with observations of the LMC). As demonstrated in \cite{PaperII}, the presence of a realistic LMC analog system is necessary to reproduce the observed anisotropy in the full-sky MW satellite population. We emphasize that our analysis conservatively marginalizes over the mass of the MW halo, which is the most important nuisance parameter governing the SHMF for halos that contain realistic LMC systems.

We combine these simulations with an empirical model for the galaxy--halo connection that is specifically developed to model dwarf satellite galaxies. This model parameterizes the slope and scatter of the relation between satellite luminosity and peak halo maximum circular velocity, the amplitude, scatter, and power-law slope of the relation between satellite size and halo size, the fraction of dark matter halos occupied by observable dwarf galaxies, and the efficiency of subhalo disruption due to the MW disk. The corresponding eight galaxy--halo connection model parameters are shown in \tabref{params}. In this work, we also incorporate the thermal-relic WDM half-mode mass, \Mhm (or the characteristic FDM SHMF mass suppression scale, $M_0$) into the model. Importantly, this empirical modeling framework allows us to parameterize and marginalize over uncertainties at the faint end of the galaxy--halo connection and the properties of the MW system in order to derive robust DM constraints.

To constrain the model using MW satellite observations, we follow the likelihood framework developed in \cite{Nadler:2018,PaperII}. This framework compares the predicted surface brightness distribution of satellites weighted by their detection probability in the relevant photometric data to the observed count, assuming a Poisson likelihood and marginalizing over the underlying Poisson rate in each surface brightness bin. We calculate detection probabilities using the state-of-the-art observational selection functions derived in \cite{PaperI}, which account for satellite detectability as a function of luminosity, size, distance, and sky position.

Although several recent studies reach similar conclusions about the galaxy--halo connection for MW satellites, our satellite model is the first to self-consistently include the population of LMC satellites. Moreover, our model comprehensively marginalizes over theoretical uncertainties in the faint-end galaxy--halo connection, including subhalo disruption efficiency and the satellite--halo size relation. In addition, the results of our MW satellite inference are consistent with predictions from hydrodynamic simulations \cite{Garrison-Kimmel180604143,Applebaum200811207}. Nonetheless, there are several modeling uncertainties in our analysis that may impact the accuracy and precision of the resulting dark matter constraints. Most importantly, the limited number of existing high-resolution cosmological simulations with realistic LMC analogs is an important problem for any analysis that attempts to fit the full-sky MW satellite population. Fitting the observed MW satellite population using zoom-in halos without realistic LMC analogs results in a severe underabundance of predicted satellites in the DES footprint due to the lack of LMC-associated satellites \cite{PaperII}. Thus, redoing our analysis without specifically-selected MW host halos would yield biased (and more stringent) dark matter constraints than those presented here.

\subsection{Fitting Procedure Details and Posterior Distributions}

Our DM limits are derived by running $10^{5}$ iterations of the MCMC sampler \texttt{emcee} \citep{emcee} to sample the eight galaxy--halo connection model parameters described in \cite{PaperII}, plus the DM model parameter of interest (i.e., $\Mhm$ for our thermal relic WDM fit and $M_0$ for our FDM fit), using $36$ walkers. The eight galaxy--halo connection model parameters are shown in \tabref{params} and described in detail by \cite{PaperII}. For both our thermal relic WDM and FDM fits, we discard a generous burn-in period of $2\times 10^4$ steps, corresponding to~$\sim 20$ autocorrelation lengths. We use the Python package \texttt{ChainConsumer}~\citep{ChainConsumer} to visualize the posterior distributions and calculate confidence intervals.

\begin{figure*}[ht]
\centering
    \includegraphics[scale=0.6]{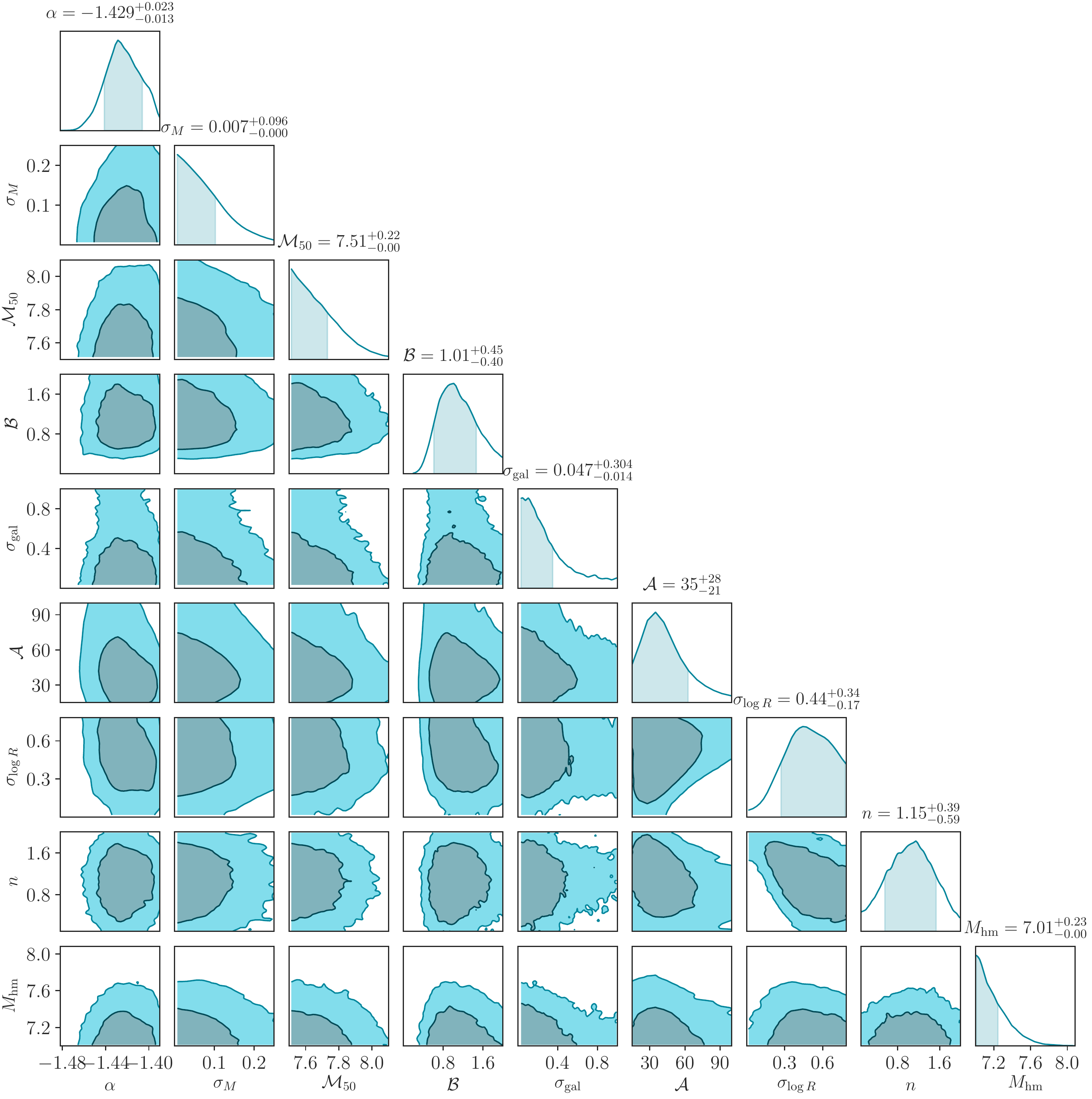}
    \caption{Posterior distribution from our fit to the DES and PS1 satellite populations for thermal relic WDM. Dark (light) shaded contours represent $68\%$ ($95\%$) confidence intervals. Shaded areas in the marginal distributions and parameter summaries correspond to $68\%$ confidence intervals. Note that $\sigma_M$, $\sigma_{\rm{gal}}$, and $\sigma_{\log R}$ are reported in $\rm{dex}$, $\mathcal{M}_{50}$ and $\Mhm$ are reported as $\log(\mathcal{M}_{50}/\Msun)$ and $\log(\Mhm/\Msun)$, $\mathcal{A}$ is reported in $\rm{pc}$, and $\alpha$, $\mathcal{B}$, and $n$ are dimensionless.}
    \label{fig:posterior_wdm}
\end{figure*}

\begin{figure*}[ht]
\centering
    \includegraphics[scale=0.6]{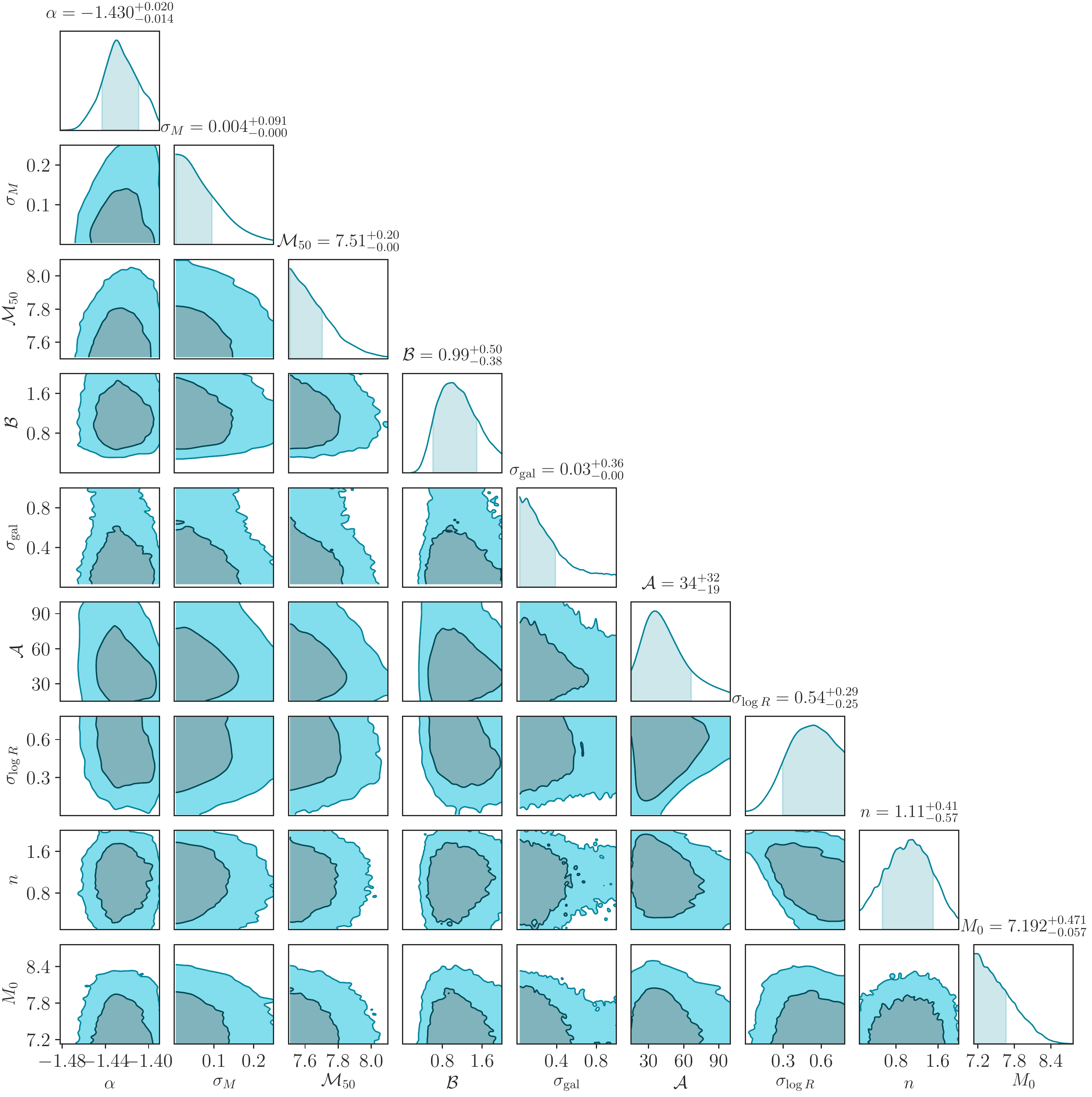}
    \caption{Same as \figref{posterior_wdm}, but for our FDM fit.}
    \label{fig:posterior_fdm}
\end{figure*}

The posterior distributions over galaxy--halo connection and DM model parameters for our thermal relic WDM and FDM analyses are shown in \figref{posterior_wdm} and \figref{posterior_fdm}, respectively. Our IDM constraints are derived using the $\Mhm$ limit from our thermal relic WDM fit; thus, we do not show a separate posterior for the IDM analysis.

\begin{table*}[t]
\begin{center}
\begin{tabular*}{0.925\textwidth}{c@{\hskip 0.35in} c@{\hskip 0.35in} c}
\hline
Parameter & Physical Interpretation & $95\%$ confidence interval \\
\hline
\hline\\[-0.5em]
$\alpha$ & Power-law slope of satellite luminosity function & $-1.46 < \alpha < -1.38$  \\
$\sigma_M$ & Scatter in satellite luminosity at fixed halo properties & $0\ \mathrm{dex}^{*} < \sigma_M < 0.2\ \rm{dex}$ \\
$\mathcal{M}_{50}$ & Peak mass at which $50\%$ of halos host galaxies & $7.5^{*} < \log(\mathcal{M}_{50}/\Msun) < 8.0$ \\
$\mathcal{B}$ & Subhalo disruption efficiency relative to FIRE simulations & $0.2 < \mathcal{B} < 1.9$ \\
$\sigma_{\mathrm{gal}}$ & Width of the galaxy occupation fraction & $0\ \mathrm{dex}^{*} < \sigma_M < 0.66\ \rm{dex}$ \\
$\mathcal{A}$ & Amplitude of relation between galaxy size and halo size & $0\pc^{*} < \mathcal{A} < 90\pc$ \\
$\sigma_{\log R}$ & Scatter in galaxy size at fixed halo properties & $0.1\ \mathrm{dex}^{*} < \sigma_M < 1.1\ \rm{dex}$ \\
$n$ & Power-law slope of relation between galaxy size and halo size & $0^* < n < 1.9$ \\
$\Mhm$ & Mass scale of thermal relic WDM SHMF suppression (\eqnref{mhm_mwdm}) & $7.0^{*} < \log(\mathcal{M}_{\mathrm{hm}}/\Msun) < 7.5$ \\
$M_{0}$ & Mass scale of FDM SHMF suppression (\eqnref{m0_fdm}) & $7.0^{*} < \log(\mathcal{M}_{0}/\Msun) < 8.1$ \\
\hline
\end{tabular*}
\end{center}
\caption{\label{tab:params}
Galaxy--halo connection and DM model parameters varied in our thermal relic WDM and FDM fits to the MW satellite population. Note that $M_0$ is constrained in a separate fit that yields similar confidence intervals for the eight galaxy--halo connection parameters. Asterisks mark prior-driven constraints. See \cite{PaperII} for details on our galaxy--halo connection model.}
\end{table*}

\subsection{Resonantly-Produced Sterile Neutrino Constraints}

To translate our upper bound on the mass of thermal relic WDM into constraints on resonantly-produced Shi--Fuller sterile neutrinos, we follow the procedure in \cite{Schneider160107553,Maamari}. 
In particular, we use the sterile neutrino transfer functions generated by \cite{Cherry:2017} using \texttt{CLASS} for a grid of sterile neutrino masses and mixing angles. 
We then compare these transfer functions to the \CHECK{$\mWDM=6.5\keV$} thermal relic transfer function that is ruled out at $95\%$ confidence by our analysis. 
We derive the limits in \figref{limits} by finding the combinations of sterile neutrino mass and mixing angle between the ``DM Underproduction'' and ``DM Overproduction'' lines in \figref{limits} that yield transfer functions which are strictly more suppressed than the ruled-out thermal relic transfer function. The overproduction (underproduction) boundaries correspond to sterile neutrino models with zero (maximal) lepton asymmetry in the Neutrino Minimal Standard Model \cite{Schneider160107553}.


We benchmark our sterile neutrino limits using the recent estimate of SHMF suppression from \cite{Lovell200301125}, which is appropriate for a $7\keV$ resonantly-produced sterile neutrino with various lepton asymmetry (or mixing angle) values. This SHMF suppression corresponds to \eqnref{wdm_shmf} with $\alpha=4.2$, $\beta=2.5$, $\gamma=-0.2$, and the relation between $\Mhm$ and lepton asymmetry $L_6$ is given for several models in \cite{Lovell200301125}. Using this SHMF in our fitting procedure, we find \CHECK{$\Mhm < 5.9\times 10^7\ \Msun$} at $95\%$ confidence. Applying linear scaling with MW halo mass to the result of the joint fit yields \CHECK{$\Mhm<7.6\times 10^7\Msun$}, which rules out the coldest sterile neutrino model presented in \cite{Lovell200301125}---corresponding to $m_s=7\keV$ and $L_6=8$---at $\gg 95\%$ confidence, consistent with our limit in \figref{limits}.



\subsection{FDM Subhalo Mass Functions}

Due to the difficulties of simulating non-linear structure formation in FDM, no consensus exists for a quantitative description of the suppression of the SHMF in this model. We therefore implemented two popular forms of the FDM SHMF to assess this uncertainty.
The nominal model described in the text is the semi-analytic model derived in \cite{Du:2019}. Our fit to this function is given by \eqnref{fdm_shmf}, which is valid for $M\gtrsim 3\times 10^8 \Msun$ and $\mFDM\gtrsim 10^{-21}\eV$ with
\begin{align}
 \tilde{\beta}(\mFDM)&=\exp\left[-\left(\frac{\mFDM}{13.7\times 10^{-22}\eV}\right)^{0.6}\right]+0.77 \\
 \tilde{\gamma}(\mFDM)&=0.22\log\left[\left(\frac{\mFDM}{10^{-22}\eV}\right)^{0.45}\right]-0.78.
\end{align}

An alternative model for the suppression of the halo mass function is derived from the ``wave dark matter'' simulations in \cite{Schive150804621}, which corresponds to \eqnref{fdm_shmf} with $\tilde{\beta} = 1.1$ and $\tilde{\gamma} = -2.2$. 
This mass function was estimated using high-redshift ($z>4$) simulation outputs and is systematically more suppressed than that derived semi-analytically in~\cite{Du:2019}. Adopting this alternative SHMF in our fitting procedure and accounting for the uncertainty in MW halo mass yields \CHECK{$M_{0}<3.4 \times 10^7\ \Msun$} at $95\%$ confidence, corresponding to \CHECK{$\mFDM>9.1\times 10^{-21}\eV$}. Thus, the current FDM SHMF uncertainty results in roughly a factor of three difference relative to our fiducial \CHECK{$\mFDM>2.9\times10^{-21}\eV$} constraint.

We caution that these uncertainties underlie FDM predictions from \emph{both} semi-analytic models and simulations. For example, \cite{Schive150804621} simulate CDM-like particles with initial conditions appropriate for FDM, and thus do not solve the Schr\"{o}dinger--Poisson system that governs FDM. This is an important caveat, because interference patterns on scales comparable to the de Broglie wavelength can potentially affect structure formation. Meanwhile, the semi-analytic treatment in \cite{Du:2019} does not explicitly account for the ``quantum pressure'' term in the Madelung transformation of the Schr\"{o}dinger--Poisson system, and makes several assumptions about the tidal evolution of subhalos with solitonic density profiles.
The derivation of robust, quantitative predictions for the FDM SHMF represents an active area of theoretical and computational study.

\end{document}